\documentclass[11pt]{article}
\newcommand{\sect}[1]{\section{#1}}
\font\uit=cmu10

\def\fraz#1#2{{\strut\displaystyle #1\over\displaystyle #2}}
\def\esp#1{e^{\displaystyle#1}}

\def\ga{\Bigl\{\Bigr.}
\def\gc{\Bigl.\Bigr\}}
\def\qa{\Bigl[\Bigr.}
\def\qc{\Bigl.\Bigr]}
\def\ta{\Bigl(\Bigr.}
\def\tc{\Bigl.\Bigr)}

\def\L{{\cal{L}}}
\def\LE{{\cal{L}}_E}
\def\HE{{\cal{H}}_E}

\def\tauf{\tau_f}
\def\taui{\tau_i}
\def\idtau{\displaystyle{\int\limits_{\tau_i}^{\tau_f}} d\tau~}

\def\auno{\alpha_1}
\def\adue{\alpha_2}

\def\sc{S_{\uit c}}
\def\xc{x_{\uit c}}
\def\xcp{\dot{x}_{\uit c}}
\def\xcpp{\ddot{x}_{\uit c}}
\def\xp{\dot x}
\def\xpdue{{\dot x}^2}

\def\ypdue{{\dot y}^2}

\def\kyn{(k\cdot y)^n}

\def\nf{n!}
\def\eu{{\epsilon}_1}
\def\ed{{\epsilon}_2}

\def\eyp{\epsilon\cdot\dot{y}}
\def\fn{f^{(n)}(\phi)}

\def\dj#1{\fraz{\delta}{\delta J_{\nu_{#1}}(\tau)}}

\def\kfi{K_{fi}}
\def\dd#1{{\cal D}[#1(\tau)]}

\date { }
\begin{document}

\title{\bf Scalar and spinning particles in a plane wave field.}

\author{A. Barducci and R.Giachetti}

\maketitle \centerline{{  Department of Physics, University of
Florence and I.N.F.N. Sezione di Firenze }}\centerline{{ Via G.
Sansone 1, I-50019 Sesto Fiorentino, Firenze , Italy\footnote{
e-address: barducci@fi.infn.it, giachetti@fi.infn.it} }}
\bigskip
\bigskip

\centerline{{ Firenze Preprint - DFF - 401/03/2003}}

\bigskip
\bigskip
\noindent

\begin{abstract}
We study the quantization problem of relativistic scalar and
spinning particles interacting with a radiation electromagnetic
field by using the path integral and the external source method.
The spin degrees of freedom are described in terms of Grassmann
variables and the Feynman kernel is obtained through functional
integration on both Bose and Fermi variables. We provide 
rigourous proof that the Feynman amplitudes are only determined by
the classical contribution and we explicitly evaluate the
propagators.
\end{abstract}
\bigskip
\bigskip

\sect{Introduction.}

Many years ago we studied the quantum mechanical interaction of a
relativistic material point (scalar particle) with an external
radiation field using the Feynman path integral method
\cite{BG}. The eigenfunctions of this problem had been
known for a long time \cite{V}, and  a
beautiful method of evaluating the Green' s functions had been
introduced by Fock and Schwinger through the solutions of the
Heisenberg quantum equations of motion in the proper time
representation \cite{FO,S}.
A direct calculation in terms of path integral
had remained unexplored probably due to mathematical difficulties. 
However, looking at the problem in a
a semiclassical way, in ref. \cite{BG} we found a canonical
transformation which made it possible to evaluate in a closed form
the Feynman kernel for a scalar particle interacting with an
external electromagnetic wave field and to obtain an exact
expression for the propagator of the theory.

Later on  a lot of interest was devoted to theories involving
anticommuting (Grassmann) variables. This was mainly raised by
supersymmetries \cite{GO}, but very soon the relevance of
anticommuting variables in many other fields was realized,
\cite{CA,BCL,BEM,GRR}. In particular it was shown that Grassmann
variables are suitable tools for giving a {\it classical}
description of spin \cite{CA,BCL,BEM} and internal degrees of
freedom of elementary particles \cite{BCL1}. These dynamical
theories described by Lagrangians involving ordinary c-numbers and
anticommuting numbers (Grassmann variables) were called {\it
pseudoclassical theories} and the general approach has been
defined as {\it pseudomechanics}, due to the special nature of the
variables occuring in the problem.

Having obtained a  pseudoclassical description of many interesting
physical systems it was very natural to investigate the
quantization of such systems by path integrals, performed both on
the ordinary and the Grassmann variables (for the general theory
of integration on Grassmann algebras see ref. \cite{BE}). This
program had already been developed in some cases; for instance it
was shown that the Wilson loop could be reconstructed as a path
integral on Grassmann variables describing the colour degrees of
freedom \cite{BCL1,IH}. Other physical systems, such as 
non-relativistic spinning particles interacting with constant electric
and magnetic field and relativistic spinning particles in external
crossed static electromagnetic wave field were studied .
In each case, the result was obtained in a quick and
straightforward way, by solving the classical equations of motion
both for  position  and  Grassmann variables. Just a bit of
caution had to be taken for systems involving an odd number of
Grassmann variables, like for  the non relativistic and for the
massive relativistic spinning particle: it was shown in ref.
\cite{BC,BBC} how to extend the path integral techniques in such
circumstances, by separating from the total phase space a coupled
one-dimensional system and studying the general solution for the
latter. For a different approach to the path integral quantization
for a non-relativistic and relativistic spinning particle by using
BRST-invariant path integral see ref. \cite{Go}.

A further powerful instrument that can be brought to bear to the
present context is the well known external source approach (in our
context real and Grassmann sources). Indeed the aim of this paper
is to provide a rigorous determination of the Feynman propagator
for a charged relativistic particle, both scalar and spinning,
interacting with an arbitrary external electromagnetic wave field
by using path integral and external sources formalism.

One can ask about the intrinsic interest of this approach apart
from the pedagogical one. We would suggest that it is at least
twofold. In the first place
the development of new techniques to solve old problems usually
increases their comprehension. Secondly, new techniques
can produce new approximation methods to apply to new problems.
In this particular case we would like to emphasize the use of these
methods in
statistical mechanics, where  an impressive number of
old problems was solved in a very fast way by using
path integrals over Grassmann variables \cite{PEV}.

The paper is organized as follows. In section 2 we present the
results for the relativistic material point in this new framework:
they are certainly simpler, due to the simpler underlying physical
model (scalar particle). Next, in section 3, we switch to the
spinning particle and we prove that this approach works well in
this case too and we evaluate the physical Feynman kernel. In the final
section 4 we express the physical kernel on a spinor basis, obtaining
a more transparent expression for the Feynman propagator.

\sect{The path integral for the scalar particle.}

The Lagrangian describing the dynamics of a relativistic scalar
particle interacting with an external electromagnetic field is
given by

\begin{equation}
\L(x,\dot x)=-m\sqrt{\xpdue}-e(\xp\cdot A)\,.
\end{equation}
This Lagrangian is singular, giving rise to the constraint

\begin{equation}
\chi=[(p-eA)^2-m^2]\,,
\end{equation}
and to a vanishing canonical Hamiltonian. According to Dirac \cite{DI} 
we define an extended Hamiltonian

\begin{equation}
\HE(x,p,\auno)=\auno[(p-eA)^2-m^2]\,,
\end{equation}
and we require the usual form of canonical Poisson brackets

\begin{equation}
\{x^\mu,p^\nu\} = -\eta^{\mu\nu}\,,~~~~~~~~\eta^{\mu\nu} =
(+,-,-,-)\,.
\end{equation}
The Lagrange multiplier $\alpha_1$ for the
 constraint $\chi$  must be chosen to be
 negative in order to have a definite positive kinetic part.
The extended Lagrangian corresponding to $\HE$ is

\begin{equation}
\LE(x,\xp,\auno)=\xpdue/(4\auno)-e(\xp\cdot A)+\auno m^2\,.
\end{equation}

We assume that the electromagnetic potential $A^\mu$ describes
the field of an external plane wave and therefore  is of
the form

\begin{equation}
A^\mu=\eu^\mu f(\phi)\,,
\end{equation}
where $\phi=(k\cdot x)$, $k^\mu=\frac 1{\sqrt{2}}(1,0,0,1)$ is the
propagation
vector and  $\eu^\mu$ a transverse real
polarization vector. We find it useful to introduce a second
transverse vector $\ed^\mu$ and the conjugate light-like vector
${\tilde k}^\mu=\frac 1{\sqrt{2}}(1,0,0,-1)$: they
satisfy the orthonormality relations
\begin{equation}
k^2={\tilde k}^2=(k\cdot\eu)=(\tilde k\cdot \eu)=(k\cdot\ed)=(\tilde
k\cdot\ed)=0\,,
\end{equation}
\begin{equation}
(\tilde k\cdot k)=-\eu^2=-\ed^2=1\,.
\end{equation}

We shall derive the amplitude for the propagation of the particle
interacting with the external field, using the path-integral
technique as developed by Feynman \cite{Feyn}. If $x_i=x(\tau_i)$
is the initial point, and $x_f=x(\tau_f)$ is the final one
$(\tau_i<\tau_f)$, we get
\begin{equation}
\kfi=K(x_f,\tauf,x_i,\taui)=c\int\limits_{(x_i,\taui)}^{(x_f,\tauf)}
\dd x \exp\ta i\idtau \LE(x,\xp,\auno)\tc
\end{equation}
where the constant c is determined by the condition
\begin{equation}
\lim_{\tauf\rightarrow\taui} K(x_f,\tauf,x_i,\taui)= \delta^4(x_f-x_i)\
\end{equation}
and the physical propagator is obtained by the integral
\begin{equation}
K_{\rm phys}=\int\limits_{-\infty}^{0}~d(\auno \Delta\tau) \kfi
\label{intalpha}
\end{equation}
where $~\Delta\tau=\tauf-\taui\,$.
The calculation is done by introducing the shift
 $x^\mu(\tau)=\xc^\mu(\tau)+y^\mu(\tau)$ where $\xc^\mu(\tau)$ is the classical path
which satisfies the equation of motion
\begin{equation}
\xcpp^\mu(\tau)=-2\auno e F^\mu_\nu(\xc(\tau)) \xcp^\nu(\tau)
\label{eqmoto}
\end{equation}
and by integrating over the deviation $y^\mu(\tau)$ from the
classical path $\xc^\mu(\tau)$ with the boundary conditions
$y^\mu(\tauf)=y^\mu(\taui) = 0 $. In equation (\ref{eqmoto})
$F^\mu_\nu$ is the electromagnetic tensor of the potential
$A^\mu$.  A series expansion in terms of $y^\mu$ then gives 
\begin{eqnarray}
{}&&\kfi=\esp{i\sc}\int\limits_{(0,\taui)}^{(0,\tauf)}\dd y
\exp\ga i\idtau\,\qa\,\fraz 1{4\alpha_1}\ypdue-e(\eu \cdot\dot
 x)\sum\limits_{n=2}^{\infty}\fraz\fn\nf \kyn\cr
{}&& \phantom{XXXXXXXX}-e(\eyp)
\sum\limits_{n=1}^{\infty}\fraz\fn\nf \kyn\qc\gc
\end{eqnarray}
where $\sc=\int_{\taui}^{\tauf} d\tau\LE(\xc,\xcp,\auno)$ is the
classical action corresponding to the extended Lagrangian $\LE$.
By introducing the functional differential operators
\begin{equation}
P_n(k,J)=\fraz 1\nf\,\prod\limits_{\ell=1}^{n}
 \ta\fraz{k_{\nu_\ell}}{i}\dj
\ell\tc \label{Pn}
\end{equation}
the propagator takes the form
\begin{eqnarray}
 {}&& \kfi=\esp{i\sc}\,\exp\ga i\idtau\qa -e(\eu\cdot \dot x)
 \sum\limits_{n=2}^{\infty}
\fn P_n(k,J)\cr
{}&&\phantom{XXXX} -e(\epsilon_{1\nu_0})
\sum\limits_{n=1}^{\infty}\fn P_n(k,J) \fraz {d}{d\tau}\,\fraz 1i\dj 0
\qc\gc K[J]\,\Bigr|_{J=0}
\end{eqnarray}
where
\begin{equation}
 K[J]=\int\limits_{(0,\taui)}^{(0,\tauf)}\dd y \exp\ga
i\idtau\,\qa\,\fraz{\ypdue(\tau)}{4\auno}+J_\mu(\tau)y^\mu(\tau)\qc
\gc \label{KJ}
\end{equation}
is the path-integral for a free system in the presence of an
external source $J^\mu(\tau)$. The latter is easily calculated and
reads
\begin{equation}
 K[J]=-i(4\pi \auno(\tauf-\taui))^{-2}
 \exp\ga \fraz{2i\auno\Delta(J)}{\tauf-\taui} \gc\,.
 \label{KJJ}
\end{equation}
Here
$\Delta(J)=\int_{\taui}^{\tauf}d\tau\int_{\taui}^{\tau}d\sigma
(\tauf-\tau)J_\mu(\tau)J^\mu(\sigma)(\sigma-\taui)$ is the
contribution of the source to the classical solution. By using the
Green function of the classical free system,
 $G(\tau,\sigma)=\qa(\tauf-\tau)(\sigma-\taui)\vartheta(\tau-\sigma)+
(\tauf-\sigma)(\tau-\taui)\vartheta(\sigma-\tau)\qc$, a simple
computation shows that
\begin{equation}
\dj i\esp{ 2i\alpha_1\Delta(J)/\Delta\tau}=(2i\alpha_1/\Delta\tau)\esp{ 
2i\alpha_1\Delta(J)/\Delta\tau}
\int_{\taui}^{\tauf}d\sigma G(\tau,\sigma) J_{\nu_i}(\sigma)
\end{equation}
and
\begin{eqnarray}
{}&&\dj i\dj 
j\esp{2i\alpha_1\Delta(J)/\Delta\tau}=(2i\alpha_1/\Delta\tau)\esp{
 2i\alpha_1\Delta(J)/\Delta\tau}\,g_{\nu_i\nu_j}\,G(\tau,\tau)+\cr
{}&&  (4i\alpha_1/\Delta\tau)\,\esp{2i\alpha_1\Delta(J)/\Delta\tau}
\int_{\taui}^{\tauf}d\sigma_1 \int_{\taui}^{\tauf}d\sigma_2
G(\tau,\sigma_1) J_{\nu_i}(\sigma_1) G(\tau,\sigma_2)
J_{\nu_j}(\sigma_2)\,.
\end{eqnarray}
Similar equations hold for the terms containing higher order
derivatives.

We therefore see that both $~~\epsilon_{1\nu_0}P_n(k,J)
\fraz {d}{d\tau}\fraz 1i\dj 0\,\exp\ga{
2i\alpha_1\Delta(J)/\Delta\tau}\gc\,\Bigr|_{J=0}~$ and $P_n(k,J) 
\,\exp\ga{
2i\alpha_1\Delta(J)/\Delta\tau}\gc\Bigr |_{J=0}\,$, as well as the 
action of all
possible mixed products of $P_n$ corresponding to different values of 
the proper time over $\exp\ga{2i\alpha_1\Delta(J)/\Delta\tau}\gc$,
are vanishing at $J^\mu=0$ due to the relations (7). In other words
\begin{eqnarray}
{}& \exp\ga i\idtau\qa -e(\eu \cdot\dot
x)\sum\limits_{n=2}^{\infty} \fn
\,P_n(k,J)\qc\gc\cdot\phantom{XXXXXXXXXXXXXXX}\cr
{}&\exp\ga
i\idtau\qa -e\sum\limits_{n=1}^{\infty} \fn\,P_n(k,J)\,
{\epsilon_{1\nu_0}}\, \fraz d{d\tau}\fraz 1i\dj 0 \qc\gc
\,K[J]\Bigr|_{J=0}=K[J]\Bigr|_{J=0}\,.
\label{PnK}
\end{eqnarray}
and we finally get
\begin{equation}
\kfi=-i(4\pi \auno(\tauf-\taui))^{-2}\,\,\esp{i\sc}\,, \label{kfi}
\end{equation}
so that the propagator is expressed in terms of the classical
action only. To evaluate the action $\sc$ we have therefore to solve
the classical equations of motion (\ref{eqmoto})
or by projecting on the
basis $k^\mu,\tilde k^\mu,\eu^\mu,\ed^\mu$ the equations
\begin{eqnarray}
{}&& (k\cdot\ddot x)=0\,,\phantom{-2 \auno e (k\cdot\dot x)f'(k\cdot
x)\,,~~~~~~}(\ed\cdot\ddot x)=0\cr
{}&& (\eu\cdot\ddot x)=-2 \auno e (k\cdot\dot x)f'(\phi)
\,,~~~~\phantom{=0} (\tilde k \cdot\ddot x)=-4\auno e(\eu\cdot\dot
x)f'(\phi)
\end{eqnarray}
where $f'(\phi)$ is the derivative with respect to the argument.
Letting $\phi_i=k\cdot x_i$ and $\phi_f=k\cdot x_f$,
their solution, although somewhat lengthy, is straightforward and
reads as follows:
\begin{eqnarray}
{}&& (k\cdot x)(\tau)=(k\cdot x_i)+\fraz{(k\cdot\Delta
  x)}{\Delta\tau}(\tau-\taui)\cr
{}&& (\eu\cdot x)(\tau)=(\eu\cdot x_i)+\qa\fraz{(\eu\cdot\Delta
x)}{\Delta\tau}+ \fraz{2\auno e}{ (k\cdot\Delta x)}\int\limits_{
\phi_i} ^{ \phi_f}d\phi f(\phi)\qc\,(\tau-\taui)
-\fraz{2\auno
 e\,\Delta\tau}
{ (k\cdot\Delta x)}\int\limits_{ \phi_i} ^{ \phi}d\phi
f(\phi)\cr
{}&& (\ed\cdot x)(\tau)=(\ed\cdot x_i)+\fraz{(\ed\cdot\Delta
x)}{\Delta\tau}(\tau-\taui)\cr
{}&&(\tilde k \cdot x)=(\tilde k \cdot x_i)+\qa \fraz{(\tilde
k\cdot\Delta x)}{\Delta\tau}+\fraz {4\auno e(\eu\cdot\Delta x)} {
(k\cdot\Delta x)^2}\int\limits_{ \phi_i} ^{ \phi_f}d\phi
f(\phi)+ \fraz {8\auno^2 e^2\Delta\tau} { (k\cdot\Delta
x)^3}\ta\int\limits_{ \phi_i} ^{ \phi_f}d\phi
f(\phi)\tc^2\cr
{}&& \phantom{XXX}-\fraz {4\auno^2 e^2\Delta\tau}{ (k\cdot\Delta
x)^2} \int\limits_{ \phi_i} ^{ \phi_f}d\phi f^2(\phi)\qc
(\tau-\taui) -\qa 4\auno e\fraz{\Delta\tau (\eu\cdot\Delta
x)}{(k\cdot \Delta x)^2} \int\limits_{\phi_i} ^{ \phi}d\phi
f(\phi)\cr
{}&& \phantom{X} + 8\auno^2 e^2\fraz{(\Delta\tau)^2}{ (k\cdot\Delta
x)^3}\,\ta\int\limits_{ \phi_i} ^{ \phi_f}d\phi
f(\phi)\tc\,\int\limits_{ \phi_i} ^{\phi}d\phi f(\phi) -
4\auno^2 e^2\fraz{(\Delta\tau)^2} { (k\cdot\Delta x)^2}\int\limits_{
\phi_i} ^{ \phi}d\phi f^2(\phi)\qc
\end{eqnarray}
where we have defined $\Delta x^\mu=(x_f^\mu-x_i^\mu)$.

The explicit form of the classical action turns out to
be
\begin{eqnarray}
{}&&\sc=\fraz{\Delta\tau}{4\auno}\qa (\fraz{\Delta
x}{\Delta\tau})^2+4\auno^2m^2\qc-\fraz{\auno e^2\Delta\tau}
{(k\cdot\Delta x)}\int\limits_{ \phi_i} ^{ \phi_f}d\phi\,
A^2(\phi)+\fraz{\auno e^2\Delta\tau} {(k\cdot\Delta
x)^2}\ta\int\limits_{ \phi_i} ^{ \phi_f}d\phi\,
A_\mu(\phi)\tc^2\cr
{}&& \phantom{XXXXXXXX} -\fraz{e\,\Delta x^\mu}{(k\cdot\Delta x)}
\int\limits_{ \phi_i} ^{ \phi_f}d\phi\, A_\mu(\phi)
\end{eqnarray}
and the final integration of (\ref{kfi}) over $\alpha_1\Delta\tau$,
as in equation(\ref{intalpha}),
gives the physical propagator in agreement with previous results
\cite{BG,S}.
\bigskip

\sect{The path integral for the spinning particle.}

We now calculate the propagator for a spinning particle in an
external plane wave. The pseudoclassical description of a spin-1/2
particle interacting with an arbitrary external electromagnetic
field has been already described in \cite{BCL} and the Lagrangian
is
\begin{equation}
\L(x,\dot x,\xi_\mu,{\dot \xi}_\mu,\xi_5,{\dot \xi}_5)=-\fraz i2
(\xi_\mu{\dot \xi}^\mu+ \xi_5{\dot \xi}_5)-\sqrt{m^2
-ieF_{\mu\nu}\xi^\mu\xi^\nu}~ \sqrt{({\dot x}^\mu-\fraz im
\xi^\mu{\dot\xi}_5)^2} -e{\dot x}_\mu A^\mu \label{spinlagrangian}
\end{equation}
Analogous results, with minor
differences, were found in \cite{BEM}, while the general theory of
quantization Fermi-Bose systems is explained in \cite{CA}.

The Lagrangian (\ref{spinlagrangian}) is singular and produces the two
first class constraints
\begin{equation}
\chi=(p-eA)^2-m^2+ie F_{\mu\nu}\xi^\mu\xi^\nu\,,
\label{constr1}
\end{equation}
\begin{equation}
\chi_D=(p-eA)\cdot \xi-i m \pi_5-(m/2) \xi_5\,,
\label{constr2}
\end{equation}
and the second class constraints
\begin{equation}
\chi_\mu=\pi_\mu - (i/2)\xi_\mu\,.
\end{equation}
The extended Hamiltonian, compatible with (\ref {constr1},
\ref{constr2}) is
\begin{equation}
\HE=\alpha_1\qa(p - eA)^2-m^2+ie F_{\mu\nu}\xi^\mu\xi^\nu\qc
+i\adue\ta(p-eA)\cdot\xi\tc+m\adue \ta\pi_5-\fraz i2\xi_5\tc
\label{secondconstr}
\end{equation}
The further second class constraints
\begin{equation}
\chi_5=\pi_5 + (i/2)\xi_5\,,
\label{chi5}
\end{equation}
will be imposed directly on the states, thereby restricting the
Hilbert space of the system. The relevant Dirac brackets are
\begin{equation}
\{\xi^\mu,\xi^\nu \}=i\eta^{\mu\nu}\,,~~~~~~~ \{\pi_5,\xi_5 \}=-1\,.
\end{equation}

According to the common practice for the path integration in 
quantum mechanics of an even number of
Grassmann variable,  \cite{BBC,F},
we substitute the $\xi^\mu$ variables with their
holomorphic combinations
\begin{equation}
\eta_1=\fraz 1{\sqrt{2}} (\xi^0+\xi^3)\,,\phantom{+i\xi2}
{\bar\eta}_1=-\fraz 1{\sqrt{2}}(\xi^0-\xi^3)\,,
\end{equation}
\begin{equation}
\eta_2=\fraz 1{\sqrt{2}}(\xi^1+i\xi^2)\,,\phantom{+\xi3}
{\bar\eta}_2=\fraz 1{\sqrt{2}}(\xi^1-i\xi^2)\,,
\end{equation}
(observe the useful identity
$z_\mu\xi^\mu=-({\bar\eta}_\alpha z_\alpha +\eta_\alpha{\bar
 z}_\alpha)$ ). For the path integration on the $(\xi_5,\pi_5)$
 variables we follow the procedure given in \cite{BC,BBC}.
 The propagator takes the form
\begin{eqnarray}
{}&K_{fi}={\displaystyle\int}
\fraz{d\pi_{5f}}i\,\,\esp{i\xi_{5f}\pi_{5f}}\, {\displaystyle
 \int\limits_{(\eta_i,x_{i}),\tau_{i}}^{({\bar\eta}_f,x_{f}),
\tau_{f}}}\,\,{\cal D}(\eta,\bar\eta)\,{\cal D}(x) {\displaystyle
\int\limits_{\xi_{5i},\tau_{i}}^{\pi_{5f}, \tau_{f}}}\,\,{\cal
D}(\xi_5,\pi_5)\,\cr
{}& \esp{[\fraz i2(\pi_{5f}\xi_5(\tauf)
+\pi_5(\taui)\xi_{5i})+\fraz 12({\bar\eta}_{\alpha
f}\eta_\alpha(\tauf) +{\bar\eta}_\alpha(\taui)\eta_{\alpha
 i})]}\,\esp{\,iS_E(\eta,\bar\eta,x)}
\cr
{}& \exp\ga{ i\displaystyle \int\limits_{\taui}^{\tauf}}\qa {\fraz
12 (\dot{\pi}_5\xi_5+\dot{\xi}_5\pi_5)-m\adue (\pi_5-\fraz
i2\xi_5)}\qc\gc
\end{eqnarray}
where
\begin{eqnarray}
{}&S_E=\int\limits_{\taui}^{\tauf} \,d\tau\,\ga \fraz
i2({\bar\eta}_\alpha
 {\dot\eta}_\alpha-{\dot{\bar\eta}}_\alpha\eta_\alpha)+
 \fraz 1{4\auno}{\dot x}^2 +\auno m^2\phantom{XXXXXXX}\cr
 {}& \phantom{XXXXXXX}+\fraz{i\adue}{2\auno}(\dot
  x\cdot\xi)-e(\dot x \cdot A)-ie\auno F_{\mu\nu}\xi^\mu \xi^\nu\gc
\end{eqnarray}
is the extended action. The shift from the classical or
pseudoclassical path
\begin{equation}
 x^\mu= x^\mu_c+ y^\mu\,,~~~~~ \eta_\alpha= \eta_{\alpha c}+
 \psi_a\,,~~~~~ \bar{\eta}_a= \bar{\eta}_{\alpha c}+
 \bar{\psi}_\alpha\,,
\end{equation}
\begin{equation}
{\xi}_{5}={\xi}_{5c}+\psi\,,~~~~~{\pi}_{5}={\pi}_{5c}+\chi\,,
\end{equation}
implies that the boundary conditions
\begin{equation}
y^\mu(\taui)=y^\mu(\tauf)=0\,,~~~~~~
\psi_\alpha(\taui)=\bar{\psi}_\alpha(\tauf)=0\,,~~~~~~
\psi(\taui)=\chi(\tauf)=0
\end{equation}
have to be imposed when integrating over the shifted variables.

We next observe that the propagator factorizes as
\begin{equation}
K_{fi}=K_5\,K_c\,K_q\,,
\end{equation}
where the meaning of the three factors will be explained here
below.

First of all it is rather evident that $K_5$ refers to the
contribution of the path integral over $\xi_5$ and $\pi_5$. The
explicit calculation has been developed in \cite{BBC} and we quote
the final result:
\begin{equation}
K_5=(\xi_{5f}-\xi_{5i})-m\adue\,\Delta\tau\,\esp{-\xi_{5f}\xi_{5i}/2}
\end{equation}
The factor $K_c$ accounts for the classical contribution including
the necessary surface terms, as explained in \cite{F},
\begin{equation}
K_c(\eta_i,{\bar\eta}_f,x_i,x_f)=\exp\ga \fraz 12(
{\bar\eta}_{\alpha f}\eta_{\alpha c}(\tauf)+{\bar\eta}_{\alpha
c}(\taui)\eta_{\alpha i}) +iS_E(\eta_c,{\bar\eta}_c,x_c)\gc
\end{equation}
and the classical variables satisfy the equations of motion
\begin{equation}
\ddot{x}_c^\mu=-2e\alpha_1
F^\mu_{\phantom{\mu}\nu}\dot{x}^\nu_c-i\adue\dot{\xi}^\mu_c-2ie
\alpha_1^2(\fraz\partial{\partial x_\mu}
F_{\nu\rho})\xi^\nu_c\xi^\rho_c\,,
\end{equation}
\begin{equation}
\dot{\xi}^\mu_c=-\fraz {\adue}
{2\alpha_1}\dot{x}^\mu_c-2e\alpha_1F^\mu_{\phantom{\mu}\nu}\xi^\nu_c\,.
\end{equation}
The third factor $K_q$ represents the contribution of the quantum
fluctuations and it is a straightforward generalization of that 
obtained in the scalar case. We can assume, without losing in 
generality, that
the plane wave field is of the form
\begin{equation}
A^\mu(x)=-2^{-1/2}(\epsilon^\mu+\epsilon^{*\mu})\,f(\phi)
=\eu^\mu f(\phi)
\end{equation}
with $\epsilon^\mu=\frac 1{\sqrt{2}}(0,-1,i,0)$ and
$\epsilon^{*\mu}=\frac 1{\sqrt{2}}(0,-1,-i,0)$. The corresponding
 electromagnetic tensor is therefore
\begin{equation}
F^{\mu\nu}(x)=f^{\mu\nu} f'(\phi)\,,~~~~~
f^{\mu\nu}=k^\mu\eu^\nu-k^\nu\eu^\mu
\end{equation}
and we finally write
\begin{eqnarray}
{}&K_q =
 {\displaystyle\int\limits_{0,\taui}^{0,\tauf}}\,\dd{\psi_\alpha(\tau),
{\bar\psi}_\alpha}\, \dd{y}~\exp\ga
~i{\displaystyle\int\limits_{\taui}^{\tauf}}\,d\tau  \ta
\qa -\fraz i2 (\psi\cdot {\dot\psi})+\fraz 1{4\auno}\,{\dot
y}^2+i\,\fraz\adue{2\auno}\, (\psi\cdot\dot y) \qc \cr
{}&-\qa e (\dot x\cdot\eu) \sum\limits_{n=2}^{\infty}\,\fraz 1\nf
 f^{(n)}(\phi)
(k\cdot y)^n + e (\dot y\cdot\eu)
\sum\limits_{n=1}^{\infty}\,\fraz 1 \nf f^{(n)}(\phi) (k\cdot y)^n
\cr
{}& + i
e\auno\,\xi^\mu\xi^\nu\,f_{\mu\nu}\,\sum\limits_{n=2}^{\infty}\,
\fraz 1\nf  f^{(n+1)}(\phi)(k\cdot y)^n + i
e\auno\,\psi^\mu\psi^\nu\,
f_{\mu\nu}\,\sum\limits_{n=0}^{\infty}\, \fraz 1\nf
f^{(n+1)}(\phi)(k\cdot y)^n\cr
{}&+
2ie\auno\,\xi^\mu\psi^\nu\,f_{\mu\nu}\,\sum\limits_{n=1}^{\infty}\,
\fraz 1\nf  f^{(n+1)}(\phi)(k\cdot y)^n\qc~ \tc \gc
\end{eqnarray}
Recalling the definition (\ref{Pn}) of the functional differential 
operators
and introducing Bose and Fermi external sources, $J^\mu$ and
$\lambda^\mu$ respectively,   we have
\begin{eqnarray}
{}&K_q= \exp\ga i{\displaystyle\int\limits_{\taui}^{\tauf}} d\tau
\qa
 i\fraz\adue{2\auno}\,\fraz
\delta{\delta\lambda_\mu}\,\fraz d{d\tau}\ta \fraz 1i \fraz
 \delta{\delta J^\mu}
\tc  -e (\dot x\cdot\eu) \sum\limits_{n=2}^{\infty}\,\fn
P_n(k,J)\cr
{}& -e \,\epsilon^{\nu_0}\,\sum\limits_{n=1}^{\infty}\,\fn \fraz
 d{d\tau} \ta
\fraz 1i \fraz \delta{\delta J^{\nu_0}}\tc
 P_n(k,J) - ie\auno \,f_{\mu\nu}\,\xi^\mu\xi^\nu
  \sum\limits_{n=1}^{\infty}\,
 f^{(n+1)}(\phi) \,P_n(k,J)\cr
 {}&-ie\auno \, f_{\mu\nu}\,\fraz\delta{\delta\lambda_\mu}
 \fraz\delta{\delta\lambda_\nu} \sum\limits_{n=0}^{\infty}\,
 f^{(n+1)}(\phi) \,P_n(k,J)
 -2ie\auno \, f_{\mu\nu}\,\fraz\delta{\delta\lambda_\mu}
 \xi^\nu \sum\limits_{n=1}^{\infty}\,
 f^{(n+1)}(\phi) \,P_n(k,J)\qc \gc \cr
 {}& \cdot\, K[J\,]\,G[\,\lambda\,]\,\Bigr |_
 {(J=0,\,\lambda=0)}
\end{eqnarray}
Here $K[J]$ is given in (\ref{KJ},\ref{KJJ}), while
\begin{eqnarray}
{}&
G[\,\lambda\,]=\displaystyle{\int\limits_{0,\taui}^{0,\tauf}}\,
{\cal D}(\psi_\alpha,\,
 {\bar\psi}_\alpha)\exp\ga
  i\displaystyle{\int\limits_{\taui}^{\tauf}}d\tau\,\qa -\fraz
   i2\,\psi_\mu{\dot\psi}^\mu -i\lambda_\mu\psi^\mu\,\qc \gc\cr
{}&=\exp\ga\,-\fraz 12
 \displaystyle{\int\limits_{-\infty}^{+\infty}}
  du\int\limits_{-\infty}^{+\infty}ds~
 \qa\, {\bar\lambda}_\alpha(u)\,D(u,s)\,\lambda_\alpha(s)\,\qc\gc\,.
\end{eqnarray}
where
\begin{equation}
D(u,s)=2\,\theta(\tauf-u)\,\theta(u-s)\,\theta(s-\taui)
\end{equation}
is the Green function of the free pseudoclassical system.

By taking into account the results obtained in equation (\ref{PnK}) for 
the scalar case,  the further relations  
\begin{equation}
f_{\mu\nu}\,(\delta/{\delta\lambda_\mu})(\delta/{\delta\lambda_\nu})=
-(\delta/{\delta{\bar\lambda}_1})\ta (\delta/{\delta{\lambda}_2})+
(\delta/{\delta{\bar\lambda}_2}) \tc
\end{equation}
\begin{equation}
 -(\delta/{\delta{\bar\lambda}_1})\ta (\delta/{\delta{\lambda}_2})+
(\delta/{\delta{\bar\lambda}_2}) \tc\,\exp\ga
 -{\bar\lambda}_\alpha\,\lambda_\alpha\gc\Bigr|_{(\lambda={\bar\lambda}
=0)}=0
\end{equation}
\noindent and
\begin{equation}
\ta \fraz\delta{\delta\lambda^\mu}\,\fraz\delta{\delta
J_\mu}\xi^\rho
f_{\rho\sigma}\fraz\delta{\delta\lambda_\sigma}\,k^\nu\fraz\delta
{\delta J^\nu} \tc\,K[J\,]\,G[\,\lambda\,]\Bigr|_{(J=\lambda=0)}
=0\,,
\end{equation}
we finally have for $K_q$ the same result as in the scalar
case, namely
\begin{equation}
 K_q=-i(4\pi\auno\Delta\tau)^{-2}\,.
\end{equation}

In conclusion, the propagator we are looking for is again fully
 determined
once we compute the classical contribution $K_c$. We find it
convenient
 to
write the latter separating in the exponent the normalization
factor of coherent states, the even Bose and even Fermi parts and
the odd Fermi
 part,
$\exp\ga {\bar\eta}_{\alpha f}{\eta}_{\alpha i}\gc$, $E_B$, $E_F$
and
 $O_F$
respectively, {\it i.e.}

\begin{equation}
K_c=\exp\ga {\bar\eta}_{\alpha f}{\eta}_{\alpha
i}+iE_B+E_F+\beta_2 O_F
 \gc\,,~~~~~(\beta_2=\alpha_2\Delta\tau)\,.
\end{equation}
$E_B$ $E_F$ and $O_F$ are obtained by solving the classical
equations of motion. Explicitly, with $\beta_1=\alpha_1\Delta\tau$,
\begin{eqnarray}
{}&E_B=\fraz{(\Delta x)^2}{4\beta_1}+\beta_1
 m^2-\fraz{e^2\beta_1}{(k\cdot\Delta
  x)}\,\displaystyle{\int\limits_{\phi_i}^{\phi_f}}d\phi
   A^2(\phi)+\fraz{e^2\beta_1}{(k\cdot\Delta x)^2}\,\ta
  \displaystyle{\int\limits_{\phi_i}^{\phi_f}}d\phi A(\phi)
     \tc^2 \cr
{}&-\fraz{e\Delta x^\mu}{(k\cdot\Delta
x)}\displaystyle{\int\limits_{\phi_i}^{\phi_f}} d\phi
A_\mu(\phi)\,,
\end{eqnarray}
\begin{equation}
E_F=-\fraz{2 e\beta_1}{(k\cdot\Delta x)} {\bar\eta}_{1
f}\,({\bar\eta}_{2 f}+\eta_{2 i})\,\ta f(\phi_f) -
 f(\phi_i)\tc\,,
\end{equation}
\begin{eqnarray}
{}&O_F=\fraz 1{2\beta_1}\,\qa {\bar\eta}_{1 f}   ({\tilde k}\cdot
\Delta x)+ {\bar\eta}_{2 f}   (\epsilon^*\cdot\Delta
x)-\eta_{1i}(k\cdot\Delta x) +\eta_{2i}(\epsilon\cdot\Delta x)
\qc\cr
{}&-e({\bar\eta}_{2 f}+\eta_{2 i})\,f(\phi_i) +\fraz e{(k\cdot
\Delta x)}\,({\bar\eta}_{2 f}+\eta_{2 i})\,
  {\displaystyle\int\limits_{\phi_i}^{\phi_f}}d\phi
f(\phi)\cr
{}& +\qa \fraz{e\Delta x^\mu}{(k\cdot\Delta
 x)^2}{\displaystyle\int\limits_{\phi_i}^{\phi_f}}d\phi
  A_\mu(\phi)+\fraz{e^2\beta_1}{(k\cdot\Delta
   x)^2}\,{\displaystyle\int\limits_{\phi_i}^{\phi_f}}d\phi
    A^2(\phi)-\fraz{2e^2\beta_1}{(k\cdot\Delta x)^3}\,\ta
   {\displaystyle\int\limits_{\phi_i}^{\phi_f}}d\phi A(\phi)
      \tc^2\cr
{}&-\fraz{2 e^2\beta_1}{(k\cdot\Delta x)^2}\,\ta f(\phi_f) +
 f(\phi_i)\tc\,{\displaystyle\int\limits_{\phi_i}^{\phi_f}}d\phi
f(\phi)+\fraz{2 e^2\beta_1}{(k\cdot\Delta x)}\,f(k\cdot
x_f)f(k\cdot
 x_i)\cr
{}&-\fraz{e \,(\epsilon^*\cdot\Delta x)}{(k\cdot\Delta
x)}\,f(k\cdot
 x_f)
-\fraz{e \,(\epsilon\cdot\Delta x)}{(k\cdot\Delta x)}\,f(k\cdot
x_i)\qc {\bar\eta}_{1f}\,.
\end{eqnarray}

We now collect all the contributions $K_5$, $K_q$, $K_c$ and we
integrate over the
Lagrange multipliers associated to the first class constraints. We
get an integrated kernel
\begin{eqnarray}
{}& K_{\rm
int}={\displaystyle\int\limits_{-\infty}^0}\,d\beta_1\,{\displaystyle
\int}\,d\beta_2 \fraz{-i\,\esp{{\bar\eta}_{\alpha f}\eta_{\alpha
i}}}{(4\pi\beta_1)^2}\, \esp{iE_B}\,\esp{E_F}\,\esp{\beta_2
O_F}\,\ta
 (\xi_{5f}-\xi_{5i})-m\beta_2\,e^{\frac 12{\displaystyle
  \xi_{5f}\xi_{5i}}}
 \tc
\cr
{}&={\displaystyle\int\limits_{-\infty}^0}\,d\beta_1{\displaystyle\int}
\,d\beta_2\,\fraz{-i\esp{{\bar\eta}_{\alpha f}\eta_{\alpha
 i}}}{(4\pi\beta_1)^2}\,\esp{iE_B}\,(1+E_F)\,\phantom{XXXXXXX}
 \cr
 {}& \phantom{XXXXXXX}\ta
  (1+\beta_2O_F)\,(\xi_{5f}-\xi_{5i})-m\beta_2\,e^{-\frac
   12{\displaystyle \xi_{5f}\xi_{5i}}}
 \tc
\end{eqnarray}

\noindent This is not yet the true physical kernel. As already
said, the second class constraint (\ref{chi5}) imposes a restriction on 
the
Hilbert space of the states. We will now produce the projection
operator that does the job. Introducing, as in
 \cite{BBC}, the
symbol $\#$ to indicate a change of sign for all the odd variables
inside the state vector, we have
\begin{eqnarray}
{}& K_{\rm phys}={\displaystyle\int}
\,\ta\,\psi_f^*(\xi_{5f})_{\rm
 phys}\,{\bar\psi}_f(\eta_f)~
 K_{\rm int} ~\psi_i({\bar\eta}_i)^\#\,\psi_i(\xi_{5i})_{\rm phys}^\#
 \,\tc\,d\mu(\eta_f)\,d\mu(\eta_i)\,d\xi_{5f}\,d\xi_{5i}\cr
{}&={\displaystyle\int}\,2^{-1/4}(1,2^{-1/2})\,\ta {\displaystyle
1\atop
 {\displaystyle\xi_{5f}}}\tc\,\langle\psi_f|\gamma_0|\eta_f\rangle
~K_{\rm int} ~\langle
-{\bar\eta}_i|\psi_i\rangle\,2^{-1/4}(1,-\xi_{5i}) \ta
{\displaystyle 1\atop {\displaystyle  2^{-1/2}}}\tc\,\cdot \cr
{}& \phantom{XXXXXXX}d\mu(\eta_f)\,d\mu(\eta_i)\,d\xi_{5f}\,d\xi_{5i}
\label{kph}
\end{eqnarray}
where $|\psi\rangle$ represents an arbitrary four component
spinor.
\bigskip

\sect{The kernel in spinor basis.}

The propagator contains the complete information on the quantum
system and in particular the wave equation itself can be deduced from
it.
This is what we are going to do in this last section and to this
purpose we begin by rewriting the different terms composing the
physical kernel as matrix elements between states of the quantum space:

\begin{equation}
K_{\rm phys}=\int\limits_{-\infty}^0 d\beta_1 \fraz
i{(4\pi\beta_1)^2}\,\esp{iE_B}\,\langle\psi_f|\gamma_0\ta {\widehat
O}_F+{\widehat{E_FO}}_F+\fraz{m\gamma_5}{\sqrt{2}}{\widehat
 E}_F+
\fraz{m\gamma_5}{\sqrt{2}}\tc |\psi_i\rangle\,.
\label{gamma5}
\end{equation}
The appearance of $\gamma_5$ in equation(\ref {gamma5}) is due to
the fact that in the basis chosen for the coherent states
$|\eta\rangle$ we have the relation
$\gamma_5|\eta\rangle=-|-\eta\rangle$ \cite{BBC}. It can be
verified by a direct calculation that the explicit form of the
operators reproducing the kernel (\ref{kph}) are the ones
given here below.
\begin{eqnarray}
{}& {\widehat O}_F=-\fraz{(\Delta x\cdot\widehat\xi)}{2\beta_1}
-\fraz {e\Delta x^\mu}{(k\cdot\Delta
 x)^2}{\displaystyle\int\limits_{\phi_i}^{\phi_f}}
 d\phi(k\cdot\widehat\xi)A_\mu(\phi) +\fraz
   {2e^2\beta_1}{(k\cdot\Delta x)^3}\ta{\displaystyle\int
   \limits_{\phi_i}^{\phi_f}}d\phi\,A_\mu(\phi)\tc^2 (k\cdot\widehat\xi)
\cr
%
{}&-\fraz{e^2\beta_1(k\cdot\widehat\xi)}{(k\cdot\Delta x)}\ta\fraz
 1{(k\cdot\Delta x)}{\displaystyle\int\limits_{\phi_i}^{\phi_f
  }}d\phi\,\ta A^2(\phi)+(A(\phi_f)\cdot A(\phi))+(A(\phi)\cdot
   A(\phi_i))\tc\,  \cr
{}&-(A(\phi_f)\cdot A(\phi_i))\tc+\fraz e{(k\cdot\Delta 
x)}{\displaystyle\int
\limits_{\phi_i}^{\phi_f}}d\phi
(A(\phi)\cdot\widehat\xi) -e(A(\phi_i)\cdot \widehat\xi)+\ta\fraz
 {e(\epsilon^*\cdot\Delta x)}{(k\cdot\Delta 
 x)}\,f(\phi_f)\phantom{XXX}\cr
  {}&+\fraz
  {e\,(\epsilon\cdot\Delta x)}{(k\cdot\Delta x)}f(\phi_i)
   \tc(k\cdot\widehat\xi)\,,
\label{OF}
\end{eqnarray}
\begin{eqnarray}
{}& {\widehat{E_FO}}_F=-\fraz e{(k\cdot\Delta x)}\qa
 -(k\cdot\widehat\xi)(\epsilon\cdot\widehat\xi)(\epsilon^*\cdot
\widehat\xi) ((\epsilon-\epsilon^*)\cdot\Delta x) \cr
{}&\phantom{XXXXXXXXXX}+(k\cdot\widehat\xi)((\epsilon+\epsilon^*)
\cdot\widehat\xi) ({\tilde k}\cdot\widehat\xi)(k\cdot\Delta
x)\qc\ta
 f(\phi_f)-f(\phi_i)\tc\,,
\label{EFOF}
\end{eqnarray}
and
\begin{equation}
{\widehat E}_F=\fraz{2e\beta_1}{(k\cdot\Delta x)}\,\ta
 (k\cdot\widehat\xi)\,
([A(\phi_f)-A(\phi_i)]\cdot\widehat\xi) \tc\,.
\label{EF}
\end{equation}

 We must now choose a representation for  the algebra of the
 operators
${\widehat\xi}^\mu$. Two possible choices are given by the
following relations:
\begin{equation}
{\widehat\xi}^\mu= i\gamma^\mu/\sqrt{2}\,,~~~~~~~~
{\widehat\xi}^\mu= \gamma_5\gamma^\mu/\sqrt{2}\,.
\label{Pauli}\end{equation}
By  substituting (\ref{Pauli}) into  equations
(\ref{OF},\ref{EFOF},\ref{EF}), the operator which represents the
physical kernel becomes
\begin{eqnarray}
{}& {\widehat K}_{\rm phys}(f,i)=\fraz
 \Gamma{\sqrt{2}}\,{\displaystyle\int\limits_0^{\infty}} ds
\fraz i{(4\pi s)^2}\,\exp\ga\, i\,\qa -\fraz {(\Delta
x)^2}{4s}-m^2s+ \fraz{e^2s}{(k\cdot\Delta
x)}\,{\displaystyle\int\limits_{\phi_i
 }^{\phi_f}}d\phi A^2(\phi)
\cr
{}&-\fraz{e^2s}{(k\cdot\Delta
 x)^2}\,\ta{\displaystyle\int\limits_{\phi_i}^{\phi_f}}d\phi
  A_\mu(\phi)\tc^2-\fraz{e\Delta x^\mu}{(k\cdot\Delta
   x)}\,{\displaystyle\int\limits_{\phi_i}^{\phi_f}}d\phi
    A_\mu(\phi)\qc\gc
\cr
{}&\cdot \qa~ m+\fraz{\gamma\cdot\Delta x}{2s}+s\,\fraz{em}{(k\cdot
\Delta x)}\, (k\cdot\gamma)\,\ta (A(\phi_f)-A(\phi_i))\cdot\gamma \tc\cr
{}& +\fraz {e}{2(k\cdot\Delta x)} \ta (k\cdot\gamma)(A(\phi_f)
\cdot\gamma) (\Delta x\cdot\gamma)-(\Delta
 x\cdot\gamma)(k\cdot\gamma)(A( \phi_i)\cdot\gamma)   \tc\cr
{}&-\fraz{e^2s}{(k\cdot\Delta x)}(k\cdot\gamma)(A(\phi_f)
\cdot\gamma)
(A( \phi_i)\cdot\gamma) +\fraz{e}{(k\cdot\Delta
x)}\,{\displaystyle\int\limits_{\phi_i
 }^{\phi_f}}d\phi (A(\phi)\cdot\gamma)
\cr
{}&\,- \fraz{e}{(k\cdot\Delta
  x)^2}(k\cdot\gamma)\, \Delta x^\mu{\displaystyle\int\limits_{\phi_i
}^{\phi_f}}d\phi A_\mu(\phi)
+\fraz{e^2s}{(k\cdot\Delta
 x)^2}(k\cdot\gamma){\displaystyle\int\limits_{\phi_i}^{\phi_f
  }}d\phi A^2(\phi)\cr
{}& \,- \fraz{2e^2s}{(k\cdot\Delta
   x)^3}\,(k\cdot\gamma)\,\ta{\displaystyle\int\limits_{\phi_i
    }^{\phi_f}}d\phi A_\mu(\phi)\tc^2 \cr
{}&\,+\fraz{e^2s}{(k\cdot\Delta
 x)^2}(k\cdot\gamma){\displaystyle\int\limits_{\phi_i}^{\phi_f
  }}d\phi\ta (A(\phi_f)\cdot\gamma)(A(\phi)\cdot\gamma)+
(A(\phi)\cdot\gamma)(A(\phi_i)\cdot\gamma) \tc\,\qc\,,
\label{kphys}\end{eqnarray}
where $\Gamma$  turns out to be a factor
 $i$ for the first representation in equations (\ref{Pauli}) and
 $\Gamma=\gamma_5$ for the second one. Notice that if we choose the 
 first and simpler
 representation we have to perform a Pauli-Gursey transformation on the
 physical spinors
\begin{equation}
|\psi\rangle \rightarrow \exp\ga i\fraz\pi 4\gamma_5\gc
|\psi\rangle \,,
\end{equation}
to obtain equation (\ref{kphys}). In this case it is then easy to
verify that
\begin{equation}
\qa\, i\gamma^\mu\fraz \partial{\partial x^\mu_f} - e\gamma^\mu
 A_\mu(x_f)+m
\,\qc\,\Delta_F(x_f,x_i|A)=i{\sqrt{2}} \, {\widehat K}_{\rm phys}\,.
\end{equation}
where we have defined
\begin{eqnarray}
{}&\Delta_F(x_f,x_i|A)={\displaystyle\int\limits_0^{\infty}} ds
\fraz {-i}{(4\pi s)^2}\,\exp\ga\, i\,\qa -\fraz{(\Delta
x)^2}{4s}-m^2s+\fraz{e^2s}{k\cdot\Delta x}
{\displaystyle\int\limits_{\phi_i}^{\phi_f}}d\phi A^2(\phi)
\cr
{}&-\fraz{e^2s}{(k\cdot\Delta
 x)^2}\,\ta{\displaystyle\int\limits_{\phi_i}^{\phi_f}}d\phi
  A_\mu(\phi)\tc^2 -\fraz{e\Delta x^\mu}{(k\cdot\Delta x)}
{\displaystyle\int\limits_{\phi_i}^{\phi_f}}d\phi
A_\mu(\phi) -\fraz{es}{2(k\cdot\Delta
x)}{\displaystyle\int\limits_{\phi_i}^{\phi_f}}d\phi\,
\sigma^{\mu\nu}F_{\mu\nu}(\phi)\qc\gc\,. \cr
\end{eqnarray}
and $\Delta_F(x_f,x_i|A)$ satisfies the squared Dirac equation
\begin{equation}
\qa\,\fraz \partial{\partial x^\mu} \fraz\partial{\partial
x_\mu}+2ie
 A^\mu
\fraz \partial{\partial x^\mu}+-e^2A^2+m^2+\fraz e 2
 \sigma^{\mu\nu}F_{\mu\nu}
\,\qc\,\Delta_F(x,y|A)=i\,\delta^4(x-y)\,.
\end{equation}
We therefore conclude that the physical kernel is indeed
the propagator for a spinor particle in the external
electromagnetic field, since it satisfies the Dirac equation
\begin{equation}
\qa\, i\gamma^\mu\fraz \partial{\partial x^\mu} - e\gamma^\mu
A_\mu(x)-m \,\qc\, \sqrt{2}\,{\widehat K}_{\rm
phys}(x,y)=-\,\delta^4(x-y)\,.
\end{equation}

\end{document}